\documentclass[aps,prb,twocolumn,floatfix]{revtex4-1}
\usepackage{graphicx,epsfig,array,color}
\usepackage{color,graphicx,amssymb,amsmath,bm}
\usepackage[utf8]{inputenc}
\usepackage[colorlinks=true,linktoc=page,linkcolor=magenta,
citecolor=cyan]{hyperref}

\newcommand{\eg}{{\it e.g. }}

\newcommand\bea{\begin{eqnarray}}
\newcommand\eea{\end{eqnarray}}
\newcommand\beq{\begin{equation}}  
\newcommand\eeq{\end{equation}}

\begin{document} 
\title{Impact of strong correlation on band topological insulator in the Lieb lattice} 

\author{Sayan Jana}
\email{sayan@iopb.res.in}
\affiliation{Institute of Physics, Sachivalaya Marg, Bhubaneswar-751005, India}
\affiliation{Homi Bhabha National Institute, Training School Complex, Anushakti Nagar, Mumbai 400085, India}
\author{Arijit Saha}
\email{arijit@iopb.res.in}
\affiliation{Institute of Physics, Sachivalaya Marg, Bhubaneswar-751005, India}
\affiliation{Homi Bhabha National Institute, Training School Complex, Anushakti Nagar, Mumbai 400085, India}
\author{Anamitra Mukherjee}
\email{anamitra@niser.ac.in}
\affiliation{School of Physical Sciences, National Institute of Science 
Education and Research, HBNI, Jatni 752050, India}

\begin{abstract}
    The Lieb lattice possesses three bands and with intrinsic spin orbit 
    coupling $\lambda$, supports topologically non-trivial band insulating 
    phases. At half filling the lower band is fully filled, while the upper 
    band is empty. The chemical 
    potential lies in the flat band (FB) located at the middle of the spectrum, 
    thereby stabilizing a flat band insulator. At this filling, we introduce 
    on-site Hubbard interaction $U$ on all sites. Within a slave rotor mean 
    field theory we show that, in spite of the singular effect of interaction 
    on the FB, the three bands remain stable up to a fairly large critical 
    correlation 
    strength ($U_{crit}$), creating a correlated flat band insulator. Beyond 
    $U_{crit}$, there is a sudden transition to a Mott insulating state, where 
    the FB is destroyed due to complete transfer of spectral weight from the FB 
    to the upper and lower bands. We show that all the correlation driven 
    insulating phases host edge modes with linearly dispersing bands along with 
    a FB passing through the Dirac point, exhibiting that the topological 
    nature of the bulk band structure remains intact in presence of strong 
    correlation. Furthermore, in the limiting case of $U$ introduced only on 
    one sublattice where $\lambda=0$, we show that the Lieb lattice can support 
    mixed edge modes containing contributions from both spinons and electrons,  
    in contrast to purely spinon edge modes arising in the topological Mott 
    insulator.
\end{abstract}

\maketitle
%----------------------------------------------------------------------
\section{Introduction}{\label{sec:I}}
%----------------------------------------------------------------------
Interplay of band theory and strong electron correlation effects has been a 
cornerstone for understanding physics of many body models and materials. 
Magnetism, superconductivity, superfluidity, all largely owe their existence to 
these two agencies.  Among these two in recent times, band theory has 
undergone a 
revolution on both theoretical~\cite{kane2005quantum, kane2005quantumZ2, 
bernevig2006quantum, zhang2009topological, moore2010birth, hasan2010colloquium, 
qi2011topological, annreview2DTI} and experimental~\cite{konig2007quantum, 
culcer2012transport, xia2009observation, annreview3DTI} fronts. Discovery of 
topologically non trivial band theory has lead to a number of breakthroughs. 
These include time reversal invariance~\cite{kane2005quantum} and crystalline 
symmetry protected band topological insulators~\cite{crys-ti, PhysRevX.7.041069}, Weyl and Dirac 
semimetals~\cite{semimetal-1,semimetal-2,semimetal-3,semimetal-4} etc.
Given this platform, it is natural to investigate the effects of correlation in 
systems that host non trivial topological bands and is being actively pursued 
theoretically~\cite{rachel2018interacting,rachel2010topological,raghu2008topological,
pesin2010mott, dzero2010topological,tran2012phase} as well as 
experimentally~\cite{shitade2009quantum,chadov2010tunable,lin2010half}.
 
In particular, we briefly discuss the study of strong local electron repulsion 
on the Kane-Mele model~\cite{rachel2010topological}. In this model, it was shown that 
the topological band insulating (TB-I) phase is stable against interaction 
effects to fairly large (non perturbative) Hubbard $U$ values. On increasing 
$U$, beyond a ($\lambda$ dependent) critical $U$, the TB-I phase evolves
into a topological Mott insulator (TM-I). Within a slave rotor mean 
field theory~\cite{florens2004slave,zhao2007self}, this Mott state is 
characterised by spin charge separation, where 
the charges are site localized while spin degrees of freedom inherit the same 
band properties as of the non interacting electrons. Hence the spinon bands, or 
equivalently a spin liquid that preserves time reversal invariance, 
exhibit non-trivial topology, thus stabilizing a fractionalized topological 
insulator (or the TM-I). 

In this spirit, in the present paper, we explore the effects of strong 
correlations on the Lieb lattice with spin orbit coupling.  Correlation driven 
magnetic, metallic and insulating phases on the Lieb lattice without spin orbit 
coupling, using Determinantal Quantum Monte Carlo method, has been recently 
reported~\cite{PhysRevB.94.155107}. 
%\textcolor{red}{
In this study, however  the role of spin orbit coupling induced  
topological phases and the impact of strong correlation driven charge fluctuations 
are in focus \cite{fn-2}. 
%} The $U=0$, 
Lieb lattice is a two dimensional (2D)
lattice with a three site primitive unit cell. For non zero spin orbit 
coupling $\lambda$, it has three bands, one of which is a flat band (FB) that is 
topologically trivial and a lower and upper bands associated with Chern numbers -1 and +1, 
implying non trivial band topology~\cite{weeks2010topological}. The lower and 
upper bands are split by $4\lambda$. With the goal to stabilize a Mott state, 
we work at half filling. At this filling, the chemical potential ($\mu$), lies 
in the flat band and the system is insulating due to the non dispersive nature 
of the FB. We term this insulator as a topological flat band insulator 
(TF-I)~\cite{fn}. The effects of disorder and interactions on FB in general has 
also been studied in recent past 
~\cite{PhysRevB.96.161104,PhysRevLett.109.096404}. In this view one of our 
motivations of focusing on the Lieb lattice 
is the existence of the flat band on which interaction effects are singular, 
due to vanishing band width. 
Given this, at the out set it would seem that a 
even tiny value of $U$ would have drastic effect on the band structure. Hence, 
will the topological flat band insulator (TF-I) 
phase be immediately destabilized by small interactions? 

The other reason 
behind choosing the Lieb lattice, is that unlike the Kane-Mele model, spin 
orbit coupling induced hopping can occur in one sublattice (indicated by 
$b-c$ in Fig.~\ref{f-1}) while, Hubbard $U$ can be applied to the other ($a$) 
sublattice. For large $U$ at half filling, the charge fluctuations on the $a$ 
sublattice will be suppressed, leaving the possibility of only spinon 
excitations. However electrons are still free on the $b-c$ sublattice.
It is known from slave rotor mean field study of ordinary Hubbard model, that 
in situations with large $U$ on one sublattice and $U=0$ or small in 
other~\cite{PhysRevLett.110.126404}, there can be delocalizing states that have 
both electronic and spinon contributions. So is there a possibility of 
realizing topologically protected edge modes with both electronic and spinon 
contributions?

We answer these questions in our work. We begin by briefly summarize our 
results. The $U=0$ state is already known to be a topological insulator. 
However, unlike in the work of Weeks and 
Franz\cite{weeks2010topological}, where a filling of 1/3 was chosen, in our 
case the chemical potential lies in the flat band. At half filling there is  
one  
particle per site on average and as mentioned above, the system is in a TF-I 
phase. Remarkably, when $U$ is switched on, in spite of the diverging ratio of 
$U$ to the (flat band) bandwidth, the lower, upper and the flat bands remain 
stable against a Mott transition. In fact the introduction of $U$, initially 
\textit{reduces}, the gap between the lower and the upper bands from the $U=0$ 
value of $4\lambda$ up to a $U_{crit}$. The $U_{crit}$ is found to increase 
with $\lambda$. We find that this phase also hosts linearly dispersing bands 
and 
a flat band at the edge, so we term it as a correlated topological flat band 
insulator (CTF-I). For $U>U_{crit}$, the gap between the lower and the upper 
bands abruptly jumps up and then keeps growing proportional to $U$. The jump is 
accompanied by complete disappearance of the flat band. This phase too is 
topological, and is analogous to the (TM-I) found in the Kane-Mele-Hubbard
model~\cite{rachel2010topological}. For all these correlation driven phases, 
the edge modes are purely made out of spinon bands. Finally in the case of $U$ 
only on the $a$ sublattice, we establish that a topological insulating phase 
can be stabilized that hosts mixed edge modes containing both electronic and 
spinon contributions.

The paper is organized as follows. In Sec.~\ref{sec:II} we discuss our model, the 
slave rotor mean field theory, and the observables. In Sec.~\ref{sec:III} we present 
our numerical results. Finally, we summarize and conclude the paper in Sec.~\ref{sec:IV}.

%----------------------------------------------------------------------
\section{Model and Method}{\label{sec:II}}
%----------------------------------------------------------------------
In this section we present our model for the Lieb lattice in presence of both spin-orbit 
coupling (SOC) and on-site Hubbard interaction term. We assume an half filled 
lattice for our calculation. Then we briefly discuss the 
slave rotor mean field theory (SRMFT) method which we employ to carry out our 
theoretical analysis. 

\begin{figure}[t]
	\centering{
		\includegraphics[width=6.5cm, height=6.5cm, clip=true]{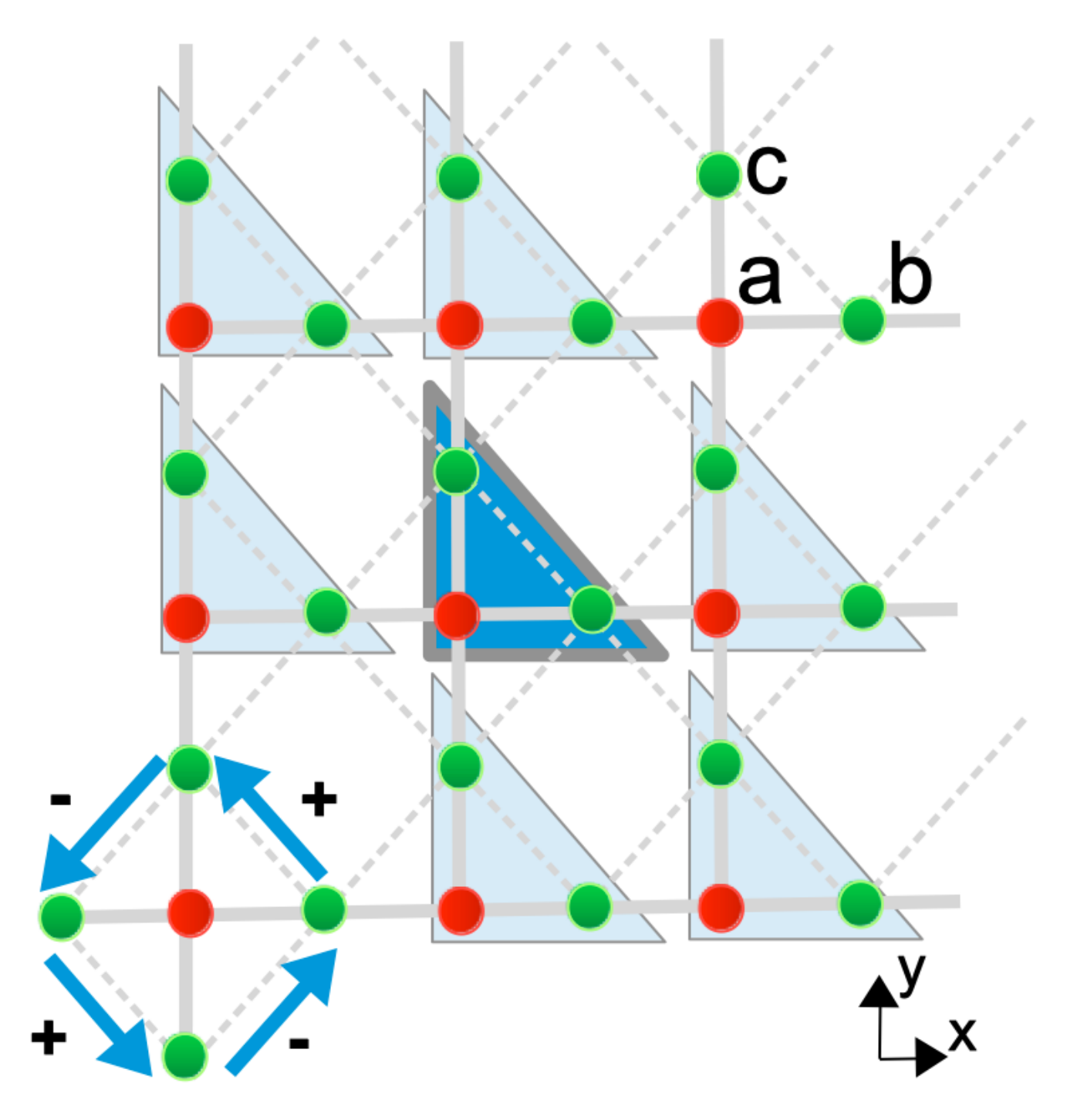}} 
	\caption{(Color online) Schematic structure of Lieb lattice is illustrated. Each 
	unit cell consists of three atoms $a$, $b$ and $c$ and are shown 
		enclosed in triangles. The solid bonds represent the nearest 
		neighbour hopping while the dashed lines represent the next nearest 
		neighbour hopping. The blue arrows show the unit vectors related to intrinsic 
		spin obit coupling and are discussed in the text.}
	\vspace{-0.0cm}
	\label{f-1}
\end{figure}

{\bf{ \textit{\underline{1. Model:}} }} In Fig.~\ref{f-1}, we present the schematic of a 
 two-dimensional (2D) Lieb lattice. The latter consists of three atoms per unit cell labelled by $a$, $b$ and $c$ as shown in 
 Fig.~\ref{f-1}, and the hopping is allowed only between nearest neighbours sites.
 The full Hamiltonian for this lattice can be written as $H=H_{t}+H_{SO}+H_{U}$. Here,
%----------------------------------------------------------------------------------------------
%--------------------------- (FIG 1
%--------------------------- 1 FIG)
%----------------------------------------------------------------------------------------------
 \begin{eqnarray}\label{e1}
&& H_{t}=-t\sum_{\langle i, j\rangle,\sigma}(c_{i,\sigma}
^{\dagger}c_{j,\sigma}+ h.c.)\ ,
\end{eqnarray}
where, $t$ is the amplitude of nearest neighbour hopping, $c_{i}^{\dagger}, (c_{i}$) represents 
the creation (annihilation) operator of electrons for the $i^{\rm th}$ lattice site with spin $\sigma$ 
and $i$ could be any of the three kinds of sites $a$, $b$ and $c$, as shown in Fig.~\ref{f-1}. The 
intrinsic spin orbit term $H_{SO}$ is defined as:
 \begin{eqnarray}\label{e2}
&&H_{SO}=i\lambda \sum_{\langle\langle ij\rangle\rangle,\sigma\sigma^{\prime}} 
({\bf{d}}^{1}_{ij}\times{\bf{d}}^{2}_{ij})\cdot{\bm{\sigma}}_{\sigma\sigma^{\prime}}
c_{i\sigma}^{\dagger}c_{j \sigma^{\prime}} \ ,
\end{eqnarray}
 Here, $\hat{\eta}_{ij}=({\bf{d}}_{ij}^1\times {\bf{d}}_{ij}^2)_{z}$=$\pm 1$ and 
$\lambda$ is the strength of the next nearest neighbor hopping induced intrinsic 
spin-orbit interaction (shown by dashed lines in Fig.~\ref{f-1}). ${\bf{d}}_{ij}^1$ and 
${\bf{d}}_{ij}^2$ are the two unit vectors %(shown by the blue arrows) 
along the two nearest neighbour bonds connecting sites $b$ and $c$ and ${\bm{\sigma}}$ 
is the vector of Pauli spin matrices. The blue arrows denote the sign of $\hat{\eta}_{ij}$
(see Fig.~\ref{f-1}).

For notational convenience, we combine $H_{t}$ and $H_{SO}$ into a single 
kinetic Hamiltonian $H_{K}$ in the following manner. We choose unit cells 
containing three sites ($a$, $b$ and $c$), shown by the triangles in Fig.~\ref{f-1}. 
We label the unit cells by $I$ and $J$. In Fig.~\ref{f-1} we indicate, by light 
colored triangles, the connection between a unit cell (shown in dark blue (dark gray)) 
and its neighbors. The hopping matrices now include both the $t$ and the spin dependent 
hopping elements. This can be written down in a compact form by simple inspection. 
 
Thus $H_{K}$ reads as follows:
 \begin{eqnarray}\label{e3}
 %\begin{split}
&& H_{K}=-\sum_{ 
I,\alpha,\sigma;J,\beta,\sigma^{\prime}}(t_{I\alpha\sigma;J\beta\sigma^{\prime}}
c_{I\alpha\sigma}^{\dagger}c_{J\beta \sigma^{\prime}} + 
h.c.)\ ,
\end{eqnarray}  
where in the $I$ and $J$ summation, $I$ can be equal 
to $J$ implying hopping within a unit cell $I$. When $I\neq J$, then the hopping 
is allowed between different unit cells $J$ that connect to $I$, as depicted in Fig.~\ref{f-1}. 

The local Hubbard interaction term can be written as, 
\begin{eqnarray}\label{e4}
H_{U}=\sum_{I \alpha}U_{\alpha}n_{I\alpha\uparrow}n_{I\alpha\downarrow}\ ,
 \end{eqnarray}
where $\alpha=a, b, c$, $U_{\alpha}$ is the strength of the on-site 
Hubbard interaction and $I$ is the unit cell index.

{\bf{\textit{\underline{2. Slave-rotor mean field theory:}}}} In order to investigate strong correlation effects, we employ a slave rotor mean field theory (SRMFT) approach. 
Below we only discuss the method briefly and refer the reader to literature 
\cite{florens2004slave,zhao2007self}  for details.

The method replaces the electron operator by a product of a bosonic degree of 
freedom (rotor) and an auxiliary fermion. The rotor is used to book keep charge 
occupations and fluctuations, while the antisymmetry of the electronic operators is preserved by the auxiliary fermion (or a spinon). 

Thus for any site in the unit cell $I$ we make the following mapping:
\begin{eqnarray}\label{e5}
&&c_{I\alpha\sigma}^{\dagger}=f_{I\alpha\sigma}^{\dagger}e^{-i\theta_{I\alpha}}\ , 
\nonumber \\
&&c_{I\alpha\sigma}=f_{I\alpha\sigma}e^{i\theta_{I\alpha}}\ .
\end{eqnarray}
where $f^\dagger_{I\alpha\sigma}$ is the spinon operator. $e^{\pm 
	i\theta_{ia}}$ represents the rotor creation and annihilation operators defined 
through its action as follows: $e^{\pm 
	i\theta_{I\alpha}}|n_{I\alpha}^{\theta}\rangle=|n_{_{I\alpha}}^{\theta}\pm1\rangle$.
Here $\alpha=a,b,c$, in the unit cell $I$.
As a standard procedure, to constrain the rotor spectrum, so that the spin and the charge 
degrees of freedom add up to physical electron occupation in the unit cell. At 
half filling, the average occupation of every unit cell is three electrons. So 
we impose the following constraint equation:
\begin{eqnarray}\label{e6}
&& \sum_{ \alpha}( 
n_{I\alpha}^{\theta}+n_{I\alpha\uparrow}^{f}+n_{I\alpha\downarrow}^{f})=3\ .
\end{eqnarray} 
with electron number equal to the spinon number 
$n_{I\alpha\sigma}^{f}=n_{I\alpha\sigma}^{e}$. 	We rewrite the original Hamiltonian 
$H=H_{K}+H_{U}$ in terms of spinon and rotor 
operators to obtain exact Hamiltonian under the slave rotor decomposition and 
then make a mean field ansatz for the full ground state:
\begin{eqnarray}\label{e7}
\lvert\Psi\rangle=\lvert\Psi^{f}\rangle \lvert\Psi^{\theta}\rangle,   
\end{eqnarray}
%--------------------------- (FIG 2
\begin{figure*}[t]
    \centering{
        \includegraphics[width=16cm, height=6.2cm, clip=true]{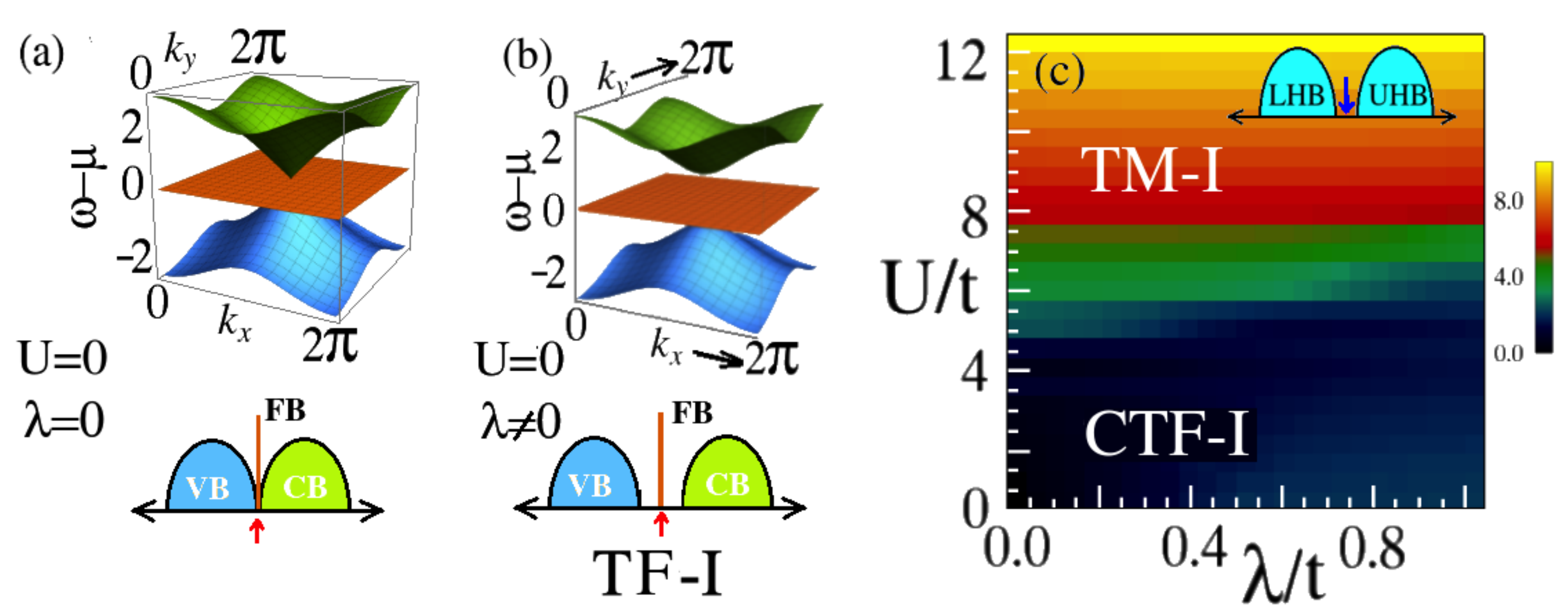}} 
    \caption{(Color online) Tight binding band spectrum of the Lieb lattice is 
        shown in panel (a), with a schematic of the density of states at the 
        bottom. (b) shows similar bands and DOS for $\lambda\neq0$. In both (a) 
        and (b) for half filling, the lower band is a filled valence band (VB), 
        the non dispersive flat band (FB) states are half filled and the upper band 
        is an empty conduction band (CB). The location of the chemical 
        potential is marked with small arrows below the  DOS. We refer to the 
        phase shown in (b) as a topological flat band insulator (TF-I).
        (c) Illustrates the $U-\lambda$ phase diagram. The background color manifests the 
        magnitude of charge gap between the lower and the upper bands. As 
        discussed in the text, the blue region corresponds to a correlated   
        topological flat band
        insulator (CTF-I) and the regime in red and green  
        belong to a topological Mott insulating phase (TM-I). The schematic DOS 
        of the TM-I is shown in the panel, where the arrow demarcates the 
        location of the chemical potential. Note that, in the TM-I phase, the 
        flat band is destroyed. For $\lambda=0$, and $U\neq0$, the system belongs to  
        topologically trivial flat band insulator and a Mott insulator, below 
        and 
        above the blue region respectively and are not shown here.}
    \vspace{-0.0cm}
    \label{f-2}
\end{figure*}
%%--------------------------- 2 FIG)
The next step is to compute $H_{f} \equiv \langle\Psi^{\theta}\lvert 
H\rvert \Psi^{\theta}\rangle$ and  $H_{\theta} \equiv\langle 
\Psi^{f}|H|\Psi^{f}\rangle$. These expressions read,
\begin{widetext}
 \begin{eqnarray}\label{e8}
H_{f}&=&-\sum_{ 
	I,\alpha,\sigma;J,\beta,\sigma^{\prime}}(\langle\Psi^{\theta}| 
	e^{-i\theta_{I\alpha}}e^{i\theta_{J\beta}}|\Psi^{\theta}\rangle
	 t_{I\alpha\sigma;J\beta\sigma^{\prime}}
f_{I\alpha\sigma}^{\dagger}f_{J\beta \sigma^{\prime}} + 
h.c.)+U/2\sum_{I,\alpha}\langle\Psi^{\theta}|
n_{I\alpha}^{\theta}(n_{I\alpha}^{\theta}-1)|\Psi^{\theta}\rangle-\mu_fN_f\ ,
\end{eqnarray}
 \begin{eqnarray}\label{e9}
 H_{\theta}&=&-\sum_{ 
	I,\alpha,\sigma;J,\beta,\sigma^{\prime}}(\langle\Psi^{f}| 
	f_{I\alpha\sigma}^{\dagger}f_{J\beta \sigma^{\prime}}
|\Psi^{f}\rangle
t_{I\alpha\sigma;J\beta\sigma^{\prime}}
e^{-i\theta_{I\alpha}}e^{i\theta_{J\beta}}+h.c.)+U/2\sum_{I\alpha}
n_{I\alpha}^{\theta}(n_{I,\alpha}^{\theta}-1)-\mu_\theta 
N_\theta\ .
\end{eqnarray}  
\end{widetext}
We thus have two coupled Hamiltonians to be solved self consistently with the 
imposition of the constraint equation (see Eq.~(\ref{e6})). We have also introduced two 
chemical potentials $\mu_f$ and $\mu_\theta$ for $H_f$ and $H_\theta$ 
respectively. These, as discussed below, will be used to satisfy the constraint 
equation on an average. Among these, Eq.~(\ref{e8}) refers to a one body problem, while
Eq.~(\ref{e9}) describes a many body rotor problem. To solve the rotor problem, we employ 
a cluster mean field description which decouples the kinetic energy term as follows.
$e^{-i\theta_{I\alpha}}e^{i\theta_{J\beta}}\rightarrow 
\langle\Psi^{\theta}| 
e^{-i\theta_{I\alpha}}|\Psi^{\theta}\rangle 
e^{i\theta_{J\beta}}\equiv\Phi_{I\alpha}
e^{i\theta_{J\beta}}$. We consider an unit cell shown in 
Fig.~\ref{f-1} as a three site cluster for which we solve the mean field 
problem. 
We assume that whatever be the value of $\Phi_{I\alpha\sigma}$, with 
$\alpha=a,b,c$, it is the same for all other unit cells according to the usual mean field 
assumption. Thus we have a three sites mean many body rotor problem to solve. 
For reducing the infinite local Hilbert space, owing to the bosonic nature of the 
rotors, we truncate the local rotor occupation to a maximum of 3. We construct 
$\langle\Psi^{\theta}| 
e^{-i\theta_{I\alpha}}e^{i\theta_{J\beta}}
|\Psi^{\theta}\rangle$ from the eigenvectors and eigenvalues of the rotor problem 
and use it to renormalize the hopping for $H_f$, before diagonalizing it. We then 
compute $\langle\Psi^{f}| 
f_{I\alpha\sigma}^{\dagger}f_{J\beta \sigma^{\prime}}
|\Psi^{f}\rangle$ and use it to solve $H_\theta$. Self consistency is terminated 
with an energy convergence criterion. At every step of the self consistency, we 
calculate $\mu_f$ and $\mu_\theta$ so that the constraint Eq.~(\ref{e6}) is 
satisfied on an average.

{\bf{\textit{\underline{3. Observables:}}}}  The main observable we focus on in 
the site projected density of states (PDOS). The PDOS is defined in general as,
$N_{\gamma}(\omega)=\sum_{\alpha,\sigma}\sum_{i_{\gamma}} 
\lvert\langle\chi_{\alpha}|i_{\gamma},\sigma\rangle\rvert^{2}\delta(\omega-
\epsilon_{\alpha})$, where, $\gamma=a, b, c$ sites in the  $I^{\rm th}$ unit cell. Here, 
$\{|\chi_{\alpha}\rangle\}$ and $\{\epsilon_{\alpha}\}$ correspond to the 
eigenvectors and eigenvalues of $H$. However, since we have split the electron 
into a rotor and a spinon at every site of our problem, we first need to reconstruct 
the (electron) single particle Green's function and then take its imaginary part to 
compute the spectral function and the PDOS. To do so, we begin with the local 
(on-site) retarded Matsubara Green's function which can be defined as
\begin{eqnarray}\label{e10}
G_{I\alpha\sigma}(i\omega_{n})&=&
-\int_{0}^{\beta} d\tau  e^{i\omega_{n}\tau} 
\langle \Psi|
c_{I\alpha,\sigma}(\tau)c_{I\alpha,\sigma}^{\dagger}(0)|\Psi\rangle \\\nonumber 
&=&-\int_{0}^{\beta}d\tau e^{i\omega_{n}\tau}\langle \Psi^f |
f_{I\alpha\sigma}(\tau)f_{I\alpha\sigma}^{\dagger}(0)|\Psi^f\rangle \\\nonumber
&&~~~~~~~~~~~~~~~\times \langle \Psi^\theta| e^{-i\theta_{I\alpha}(\tau)} 
e^{i\theta_{I\alpha}(0)}|\Psi^\theta\rangle\ .
\end{eqnarray}
The above decomposition of electron Green's function into a convolution of rotor 
and spinon Green's functions is possible for the chosen mean field ansatz 
$|\Psi^{f}\rangle |\Psi^{\theta}\rangle$.
The spinon correlator in Eq.~(\ref{e10}) can be calculated as
\begin{eqnarray}\label{e11}
&&\frac{1}{2}\sum_{\sigma}\langle 
f_{I\alpha\sigma}(\tau)f_{I\alpha\sigma}^{\dagger}(0)\rangle \nonumber \\
&&=\frac{1}{2}\sum_{\alpha\sigma}\lvert \langle 
\chi^f_{\alpha}|I\alpha,\sigma\rangle 
\rvert 
^{2} [1-n_{f}(\epsilon^f_{\alpha}-\mu_{f})]e^{-\tau(\epsilon^f_{\alpha}-\mu_{f})}\ .
\label{SC}
\end{eqnarray}
Here, $\{|\chi^f_\alpha\rangle\}$ and $\{\epsilon_{\alpha}^f\}$ are the spinon 
eigenvectors and eigenvalues respectively. The rotor correlator in Eq.(\ref{e10}) 
can be expressed as 
\begin{eqnarray}\label{e12}
&&\langle e^{-i\theta_{I\alpha,\sigma}(\tau)}e^{i\theta_{I\alpha,\sigma}(0)}\rangle 
\nonumber \\
&&= \frac{1}{Z_{\theta}}\sum_{m,n}e^{-\beta\epsilon_{m}}\langle m\lvert 
e^{-i\theta_{I\alpha,\sigma}}\rvert n \rangle \langle n \rvert 
e^{i\theta_{I\alpha,\sigma}} \rvert m\rangle 
e^{\tau(\epsilon_{m}-\epsilon_{n})}\ .
\end{eqnarray}
where, $\{\epsilon_{m}\}$ and $\{\rvert m\rangle\} $ are the eigenvalues and 
corresponding eigenvectors of the rotor Hamiltonian $H_{\theta}$. 
Here, $Z_{\theta}$ is the rotor partition function defined as 
$\sum_{m}e^{-\beta\epsilon_{m}}$. Using Eq.(\ref{e10}), the integration over 
imaginary time $\tau$ can be performed. We then analytically continue back to 
the real frequency to obtain $G_{I\alpha\sigma}(\omega)$. The PDOS is obtained 
from it's imaginary part as usual.

We solve effective tight binding models based on the 
slave rotor calculations, to characterize the band topologies. We discuss this 
later in the text.

%----------------------------------------------------------------------
\section{Results}{\label{sec:III}}
%----------------------------------------------------------------------
We begin with the evolution of the charge gap with $U$ and $\lambda$ and then 
discuss the topological properties of the insulating states. Here we assume $U$ 
to be same on all three sites in the unit cell and the filling of 1/2, or three 
electrons per unit cell. 

%--------------------------- (FIG 3
\begin{figure}[t]
    \centering{
        \includegraphics[width=8.5cm, height=8.0cm, clip=true]{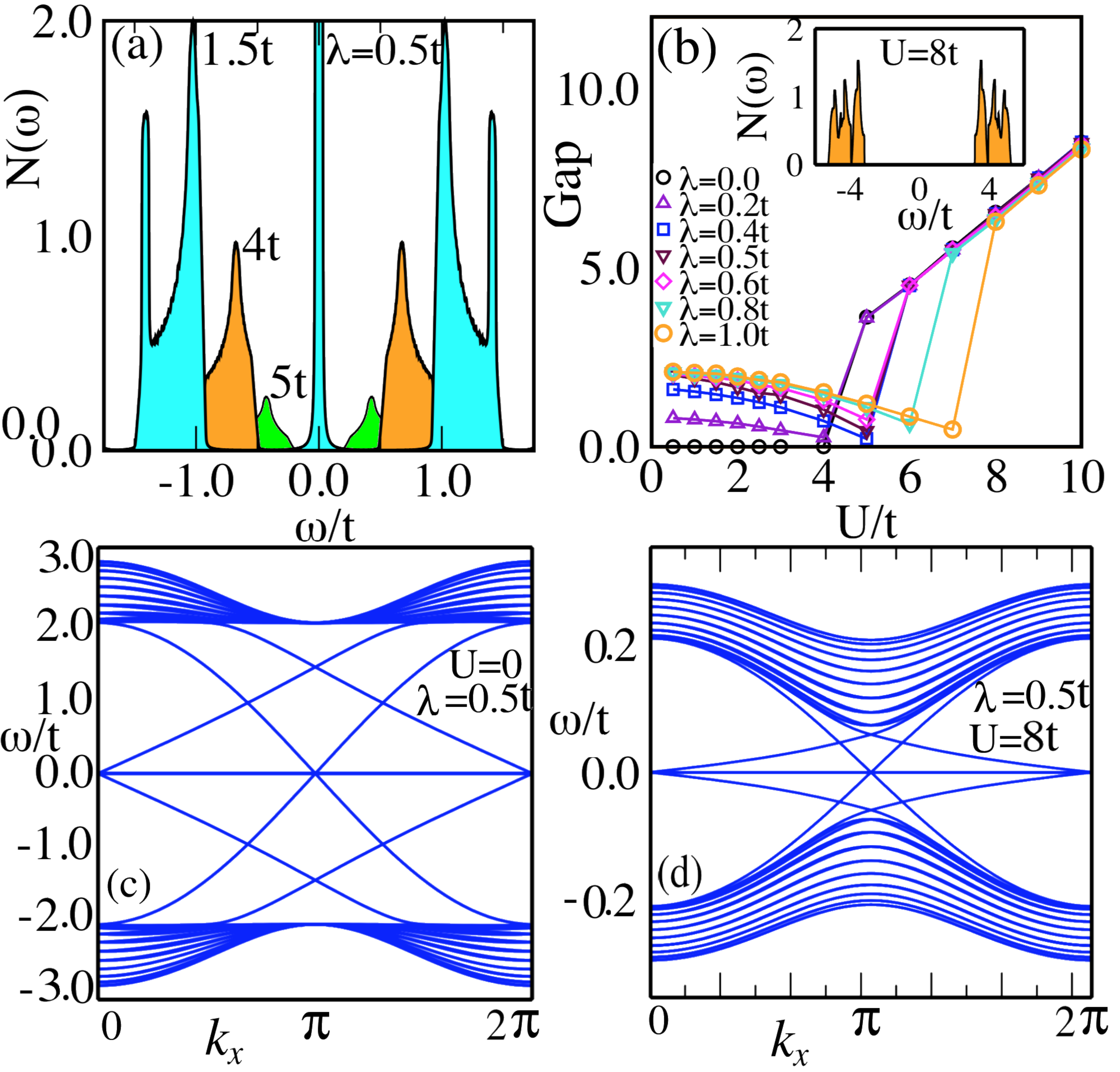}} 
    \caption{(Color online) (a) Total density of states (DOS) for three $U$ values at fixed 
        $\lambda=0.5t$ is shown. We see that the gap reduces in magnitude with 
        increasing 
        $U$, up to 5$t$. For larger $U$, the system develops gap that scales 
        linearly 
        with $U$, as shown for $U=8t$ and $\lambda=0.5t$ in the inset of panel 
        (b). 
        The main panel (b), exhibits the evolution of the gap between the lower 
        and 
        upper bands with $U$ and for a fixed value of $\lambda$. Panels (c) and 
        (d) 
        depict the edge modes of the full Hamiltonian in the non interacting case 
        (c) 
        and for the spinon Hamiltonian ($H_f$) in the panel (d). 
        See text for discussion.}
    \vspace{-0.0cm}
    \label{f-3}
\end{figure}
%--------------------------- 3 FIG)
%{\bf{\textit{1. Correlation effects:}}} 
{\bf{\textit{\underline{1. Correlation effects:}}}}
The $U=0$ and $\lambda=0$ bands are shown in 
Fig.~\ref{f-2}(a) and the schematic DOS is shown below it. For the filling considered here, the lower 
valence band (VB) is completely filled and the flat band 
FB is partially filled. Since the FB is dispersionless, the electrons in this 
band are localized due to destructive quantum interference. Thus although, the 
empty upper conduction band (CB) touches the FB and the filled VB, the system is a 
topologically trivial \textit{gapless} flat band insulator. For $U=0$, but non 
zero 
$\lambda$, a band gap opens up 
between the VB, FB and CB. We note that, such gap would open up even if the signs 
of direction dependent spin orbit hopping were all same, however, the bands 
would be topologically trivial with zero Chern number. Nonetheless, since our focus is on the 
interplay of topology and correlation effects, we choose the 
signs such that it introduces a chirality around the $a$ site sublattice. The 
band splitting and the schematic DOS are shown in Fig.~\ref{f-2}(b). We refer 
to this phase as TF-I, as defined in the introduction. In the $U-\lambda$ phase 
diagram shown in  Fig.~\ref{f-2}(c), we observe that this split 
band scenario along with the FB, survives even in the presence of correlations 
up to a critical $U$. The dependence of $U_{crit}$ on $\lambda$ is demarcated
by the boundary between the blue and green region. The phase below 
the $U_{crit}$ is still an insulator in the sense discussed above and 
we call this phase as CTF-I. Just above the blue region, for strong $U$, the 
gap between 
the lower and upper bands jump discontinuously (within numerical resolution) 
and the FB disappears. 
%\textcolor{red}{
The stability of the FB at finite $U$ in the limit of vanishing one electron bandwidth has been reported in Hartree-Fock calculations \cite{GOUVEIA2016292} and has been discussed in literature \cite{PhysRevLett.109.096404}. It can be  understood as follows. The flat band implies non dispersive electronic states that are described by real space (real valued) wave-functions that have large degeneracy. On including $U$, the wave-functions adjust by reducing the real space overlaps. However, beyond $U_{crit}$, even small wave function overlaps are too costly and the spectral weight is transferred to the lower and upper Hubbard sub bands.%} 

As mentioned above, we do not find this transfer to be gradual, rather a sudden 
change at a $\lambda$ dependent critical correlation strength ($U_{crit}$). We 
refer to this phase as TM-I, where the VB and the CB change into the lower and 
upper Hubbard sub-bands, namely LHB and UHB respectively. The schematic 
DOS for this case is shown in the phase diagram. We will discuss the determination 
of the topological nature of the bands for all phases later in the paper.

We would like to stress on two important issues with regards to stability of 
the correlated phases. First, in a slave rotor mean field theory, one drops 
the U(1) gauge fluctuations~\cite{PhysRevLett.95.036403}. These fluctuations 
are however not negligible in two dimensions. Hence, for the stability of the mean 
field phases, we assume that there are layers of such Lieb lattices weakly 
coupled to each other. Thus, our mean field results should be thought of as valid  for quasi 
2D Lieb lattice. 
%\textcolor{red}{
The second important issue is that at large 
enough $U$, as in the Kane-Mele-Hubbard model, the TM-I is likely to be 
unstable towards magnetic phases. This is particularly important in the present case because one expects flat band ferromagnetism to appear~\cite{Katsura_2010} for any value of $U$ at zero temperature. These cannot be captured within the SRMFT as we are in the paramagnetic regime. In the present paper, our goal is to simply study the effect of $U$ on uncorrelated topologically non-trivial bands. 
Also being a two dimensional lattice, the $T=0$ long range magnetic order is likely to get suppressed at any finite temperature due to Mermin-Wagner theorem~\cite{PhysRevLett.17.1133}.%} ~ 
~Capturing magnetism within a strong correlation 
slave boson theory with Kotliar-Ruckenstein representation~\cite{PhysRevLett.57.1362} at $T=0$ is currently being worked out by the authors. It will also be of interest 
to see if there is a metallic phase close to the $U_{crit}$ particularly at 
very small $\lambda$. Such kind of metallic phase has been theoretically 
predicted to exist between correlated band and Mott insulator in the ionic 
Hubbard model\cite{PhysRevLett.97.046403} and for the pyrochlore 
lattice~\cite{pesin2010mott}.

We now briefly discuss the various indicators used to construct the $U-\lambda$ 
phase diagram. In Fig.~\ref{f-3}(a), we show the DOS for three $U$ 
values for $\lambda=0.5t$. For $\lambda=0.5t$ and $U=0$, the gap between the lower and 
upper bands is $4\lvert\lambda\rvert(=2t)$, not shown. This gap gradually decreases with 
increasing $U$ up to 
$U=5t$. However the overall feature of lower and upper bands and the FB 
at $\omega=\mu$ survives. At $U=6t$, the gap jumps suddenly, as can be seen in Fig.~\ref{f-3}(b)
that manifests the evolution of the gap between the upper and the lower bands  
as a function of $U$ for different $\lambda$ values. At $U=6t$, concomitant with the sudden 
increase of the gap, the FB disappears. As seen in (b), on further increase of $U$, the 
gap grows linearly with 
$U$. The inset in (b), shows the DOS at $U=8t$, which has a gap of about 
$\sim8t$ as in a Mott insulator with no FB contribution from the bulk band. In panel 
(b) we also see that the suppression of the gap with $U$ is a common feature at 
all $\lambda$ values explored. Unfortunately, within our numerical resolution, 
we cannot capture it for very small $\lambda$, where the gap is indeed closed.
%--------------------------- (FIG 4
\begin{figure}[t]
    \centering{
        \includegraphics[width=8.6cm, height=8.0cm, clip=true]{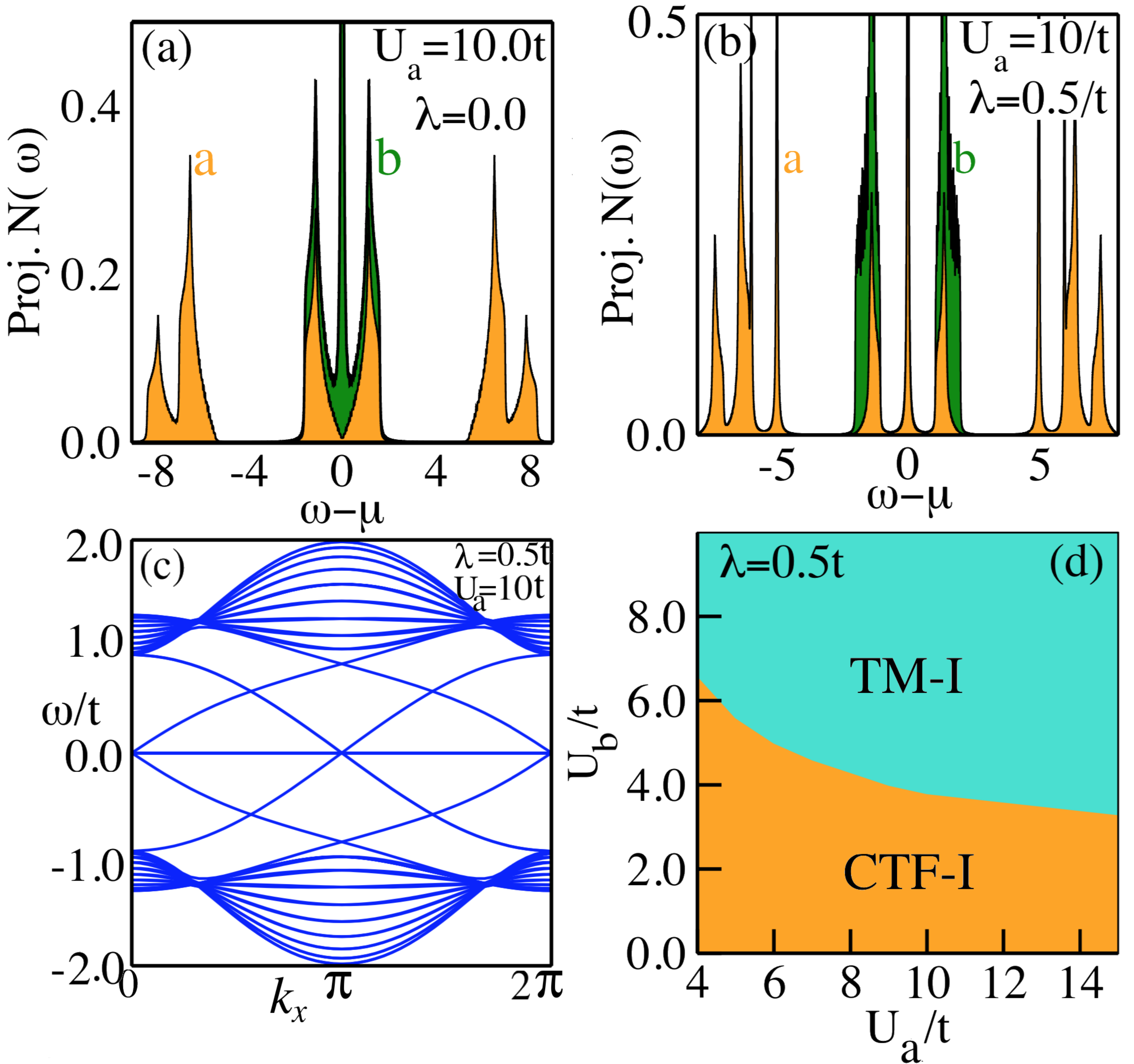}} 
    \caption{(Color online) Projected DOS on the $a$ and $b$ sub-lattices for 
        $U_a=10t$, and $\lambda=0$ (a) and $\lambda=0.5t$ (b) respectively. The labels ``$a$"
        and 
        ``$b$" in the panels indicate the projected sub-lattices. The DOS projected 
        on 
        the $c$ sub-lattice is identical to $b$ and is not shown here. For 
        $\lambda=0$, 
        low energy bands from $a$ and $b$ sublattice overlap at $\omega=\mu$, 
        apart 
        form the FB. This state is thus a metal. For $\lambda=0.5t$, we see that 
        spectral weight is pushed away from $\omega=\mu$, leaving only the 
        non-dispersive FB at the chemical potential. Panel (c), shows the edge 
        modes for the spinon-electron Hamiltonian ($H_{fc}$) for $U=10t$ and 
        $\lambda=0.5t$. 
        The feature of it is discussed in detail in the main text. Panel (d) shows $U_a-U_b$ 
        phase 
        diagram for $\lambda=0.5t$. We find two phases which are a correlated 
        topological flat band 
        insulator 
        and a topological Mott insulator. Panel (c) shows that the CTF-I phase hosts 
        non trivial topological edge modes.}
    \vspace{-0.0cm}
    \label{f-4}
\end{figure}
%--------------------------- 4 FIG)

{\bf{\textit{\underline{2. Band topology:}}}} In Figs.~\ref{f-3}(c) and (d) we show the edge modes 
computed considering the slab geometry, with periodic boundary condition along the $x$ direction and 
open boundary condition in the transverse ($y$) direction with a strip of $N_{y}=10$ unit cells. 
The bands are plotted with $k_x$, while the edge exists along $y$. The edge terminations are 
$a-b$ at the lower edge and $c$ at the upper edge. This termination is chosen to keep the results consistent 
with the three sites unit cell shown in Fig.~\ref{f-1}.
For $U=0$ and finite $\lambda=0.5t$, as is well known, we find linearly 
dispersing edge modes crossing the chemical potential ($\mu$) at a single Dirac 
point apart from the non dispersive FB contribution. For large $U=8t$, we show 
the bands for 
the spinon Hamiltonian, whose hopping element between sites $i$ and $j$ is 
renormalized by the $\langle\Psi^{\theta}| 
e^{-i\theta_{I\alpha\sigma}}e^{i\theta_{J\beta\sigma^{\prime}}}|\Psi^{\theta}
\rangle$. We note that even in the Mott phase the virtual charge fluctuation 
within the cluster (used to solve the cluster mean field theory for 
$H_{\theta}$), allows a spinon hopping 
term in $H_f$. We see that apart from the bandwidth 
renormalizations, the spinon bands also support the linearly dispersing edge 
modes meeting at the Dirac point and the flat band. Thus the TM-I phase hosts
purely spinon edge modes with localized charges.

{\bf{\textit{\underline{3. Electron-spinon edge modes:}}}} We now consider the case where $U_a\neq 
U_b=U_c$. The reason this case is interesting, is because in the specific case 
of $U_b=U_c=0$, the slave rotor 
decomposition is performed only on the $a$ sites. Due to this, if a band or  
Mott insulating state is stabilized, the edge modes will have contributions 
arising from both spinons and electrons. With this idea in mind, we repeat the 
above 
analysis for fixed $U_a$ and different values of $\lambda$. The details are 
presented in the Appendix.
Figs.~\ref{f-4} (a) and (b) depict the projected DOS of the $a$ and $b$ 
sub-lattices for $U_a=10t$ at $\lambda=0$ and $\lambda=0.5t$, respectively. 
The PDOS for the $c$ sublattice is identical to that of the $b$ sublattice and is 
not shown here. We find that for zero spin orbit coupling, the system behaves 
as a metal. 
In panel (a), there are two high and low energy bands with band edges at +5$t$ 
and -5$t$ respectively due to the presence of large $U$ and are related to the 
lower and upper Hubbard sub-bands. However, because of the no correlation 
strength on the $b$ and $c$ 
sub-lattices, there are two other bands, below and above the Fermi energy, that 
have overlap at the chemical potential. In addition there is the FB pinned at 
$\omega=\mu$. From the partial DOS we see that the overlap at $\mu$ comes from 
both the $a$ and $b$ (and $c$) sub-lattices. This metal is clearly a strongly 
correlated metal. Incorporating finite $\lambda$, here shown for 
$\lambda=0.5t$, from panel (b) we see that, while the high energy upper and the 
lower sub-bands maintain their positions, spectral weight is transferred away 
from $\omega=\mu$ to the low energy sub-bands. 

The only contribution at $\omega=\mu$, is now arising from the FB. This is a 
new insulating phase, where the gap between the VB and CB is about $\sim 
4\lambda$ and it thus primarily 
controlled by the spin orbit coupling. Hence, this phase is a spin orbit 
coupling 
driven flat band insulator arising out of a strongly correlated metal. 
While, as discussed below, this too is a topological flat band insulator, it 
differs from the earlier CTF-I, in that there are extra sub-bands. In 
Fig.~\ref{f-4}(c), 
we show the edge mode in the strip geometry with the same edge termination as in 
panel (c) and (d) of Fig.~\ref{f-3}. The calculation is carried out as before. 
The only difference is that instead of a purely spinon Hamiltonian ($H_f$) as 
in the earlier case, the Hamiltonian ($H_{fc}$) contains 
both spinon and electron operators (see Appendix for details). In panel (c) we 
find that the linearly dispersing edge modes persist in this new 
CTF-I phase. However, because $H_{fc}$ 
involves both electrons and spinons, the \textit{`electron-spinon edge mode'} 
wave function will have contributions arising from both electrons and spinons. Finally, 
in Fig.~\ref{f-4}(d) we show the $U_a-U_b$ phase diagram at $\lambda=0.5t$. We find 
that for a fixed $U_a$, the system is a CTF-I up to a 
critical value of $U_b(=U_c)$. The critical value reduces with increasing 
$U_a$. Above this critical value, the system reaches to a TM-I phase. 
The entire insulating phase space has bands with non trivial band 
topology (Chern number=$\pm 1$).

%----------------------------------------------------------------------
\section{Summary and Conclusions}{\label{sec:IV}}
%----------------------------------------------------------------------
To summarize, in this article we have investigated the interplay between strong 
correlation and intrinsic spin orbit coupling effects in the Lieb lattice. 
%\textcolor{red}{
Since our focus has been to study the impact of correlation driven charge fluctuations on nontrivial band topology, we have worked in a paramagnetic regime explicitly~\cite{fn-2}. Similar study on the 
Kane-Mele-Hubbard model has shown that the topological Mott state captured within slave rotor mean field theory qualitatively agrees with cluster dynamical mean field theory~\cite{cdmft}, 
determinantal~\cite{rachel2010topological} and variational~\cite{vqmc} quantum Monte Carlo method. We believe that this justifies the use of SRMFT in the present case of Lieb lattice.%}

Due to the presence of three lattice sites $a$, $b$ and $c$ in a unit cell of 
Lieb lattice (see Fig.~\ref{f-1}) and the existence of the FB in the spectrum, the 
$U-\lambda$ phase diagram 
is richer in this case compared to the Kane-Mele-Hubbard 
model~\cite{rachel2010topological} based on 
hexagonal lattice structure with two sites ($a-b$) unit cell. Moreover, the 
freedom of incorporating Hubbard $U$ 
on either $a$ sub lattice or all three sites ($a$, $b$, $c$) in the unit cell of 
Lieb lattice enables us to explore 
the possibility of obtaining new exotic phases with topological character. 

It is indeed interesting that in spite of singular effects of correlation on the flat band, the FB survives to fairly large correlation effects. When we 
allow correlation strength to be same on all three sites in the unit cell, we obtain 
a topological flat band insulator (at $U=0$), then a correlated topological 
flat band insulator and finally a topological Mott insulator, with increasing 
$U$. All the phases exhibit linearly dispersing and flat band contributions to the spinon edge 
modes and hence show signatures of topologically non trivial bulk bands. 
Further, when correlation is allowed only on the $a$ sublattice, we find a 
correlated metal where 
all three sub-lattices participate in the conduction at 
$\lambda=0$. For finite $\lambda$, the spectral weight are pushed away from the 
flat band, leading again to a CTF-I. The corresponding edge modes also exhibit 
linear band crossing. 
However, in sharp contrast to the previous case, here the edge modes contain 
contributions from electronic degrees of freedom residing on the $b$ and $c$ sub-lattices 
and the spinon modes from the $a$ sub lattice. This kind of `mixed' edge modes are novel
and likely to have transport signatures distinct from either purely electronic 
or purely spinon edge modes. As far as practical realization of our geometry is 
concerned, Lieb lattice has 
been realized recently in optical lattice 
systems~\cite{mukherjee2015observation, vicencio2015observation}. In such 
systems one can control the hopping parameter $t$ and repulsive local  
interaction strength $U$, and thus one can realize the repulsive 
Hubbard model in those systems ~\cite{bloch2008many}. Also one can engineer the effect of spin-orbit coupling $\lambda$
in optical lattice systems~\cite{zhang2015spin}. For \eg Mott insulator and topological Haldane model have been realized
experimentally in optical lattice systems~\cite{jordens2008mott, 
jotzu2014experimental}. Based on this, it is highly likely to realize our 
theoretical $U-\lambda$ (interplay between strong correlation and band 
topology) phase diagram of Lieb lattice, engineered in optical lattice systems.
\vskip -1cm
\begin{center}
    \textbf{ACKNOWLEDGEMENTS}
\end{center}
We acknowledge S. D. Mahanti, Michigan State, and Kush Saha, NISER, for 
stimulating and useful discussions.

\begin{center}
    \textbf{APPENDIX}
\end{center}
%\vskip +0.1cm 
 %----------------------------------------------------------------------------
\begin{center}
    \textbf{Slave rotor calculation for $U$ only on the $a$ sublattice}
\end{center}
%-------------------------------------------------------------------------------
Here we briefly outline the details of the slave rotor calculation for the case 
where $U$ is implemented only on the $a$ sublattice. The main difference from 
what is discussed in the methods section (see Sec.~\ref{sec:II}) of the paper is that 
the slave rotor decomposition is only performed on the $a$ sites of the Lieb lattice and 
the constraint is also imposed on the $a$ sites. This is the standard approach 
used in SRMFT when there there are sites with and 
without $U$, or there are sites with large and small 
$U$~\cite{PhysRevLett.110.126404}. Thus, there are electronic, rotor 
and spinon operators in this case. We chose the following product ansatz, 
\begin{eqnarray}\label{a1}
\lvert\Psi\rangle=\lvert\Psi^{fc}\rangle \lvert\Psi^{\theta}\rangle\ ,   
\end{eqnarray}
Here $|\Psi^{fc}\rangle$, refers to spinon-electron wavefunction.
Then following the same procedure as described in the methods section, we 
obtain the 
following coupled Hamiltonians, which has been solved self-consistently in a manner 
similar to that discussed in Sec.~\ref{sec:II}. The two coupled Hamiltonians 
can be written as follows:

\begin{widetext}
    \begin{eqnarray}\label{a2}
H_{fc}&=&-\sum_{ 
    I,J,\beta,\sigma}(\langle\Psi^{\theta}| 
e^{-i\theta_{Ia}}|\Psi^{\theta}\rangle
t_{Ia\sigma;J\beta\sigma}
f_{Ia\sigma}^{\dagger}c_{J\beta \sigma} +
h.c.)+H_{SO}+U/2\sum_{I}\langle\Psi^{\theta}|
n_{Ia}^{\theta}(n_{Ia}^{\theta}-1)|\Psi^{\theta}\rangle-\mu_fN_f\ \ ,
\end{eqnarray}
\begin{eqnarray}\label{a3}
H_{\theta}&=&-\sum_{ 
    I,J,\beta,\sigma}(\langle\Psi^{fc}| 
    f_{Ia\sigma}^{\dagger}c_{J\beta \sigma}
    |\Psi^{fc}\rangle
    t_{Ia\sigma;J\beta\sigma}
    e^{-i\theta_{Ia}}+h.c.)
    +\langle\Psi^{fc}|H_{SO}|\Psi^{fc}\rangle+U/2\sum_{I}
    n_{Ia}^{\theta}(n_{I,a}^{\theta}-1)-\mu_\theta 
    N_\theta\ \ ,
    \end{eqnarray}  
\end{widetext}
Here the summation over $\beta$ runs over $b$ and $c$ and $I$,$J$, run over the unit cells as 
before. $H_{SO}$ is the hopping 
Hamiltonian containing only electron operators, given in Eq.~(\ref{e2}). This 
term operates only on the $b-c$ sub-lattice. The 
constraint equation employed is: 
\begin{eqnarray}\label{a4}
&& 
n_{Ia}^{\theta}+n_{Ia\uparrow}^{f}+n_{Ia\downarrow}^{f}=1\ .
\end{eqnarray} 
Note that, in solving the rotor Hamiltonian (Eq.~(\ref{a3})), we do not need to 
perform the kinetic term mean field decoupling as rotor operators contain only quadratic terms.

%%%%%%%%%%%%%%%%%%%%%%%%%%%%%%%%%%%%%%%%%%%%%%%%%%%%%%

%-----------------------------------------------

%-----------------------------------------------
\bibliography{bibfile}{}

\begin{thebibliography}{50}
\bibitem [{fn-()}]{fn-2}%
  \BibitemOpen
  \href@noop {} {\ }\bibinfo {note} {Given that we study the system at half
  filling, the Lieb lattice is known to host flat-band ferromagnetism in the
  ground state at any $U>0$. Nonetheless, in the present paper we study the
  model in the paramagnetic regime to study the interplay of charge
  fluctuations and band topology.}\BibitemShut {Stop}%
\bibitem [{fn()}]{fn}%
  \BibitemOpen
  \href@noop {} {\ }\bibinfo {note} {For half filling, our chemical potential
  lies in the topologically trivial flat band (FB). Nonetheless, the system
  still possesses two other bands with non-trivial topology. By `topological'
  flat band insulator, we imply that the bulk band has non-trivial topology,
  while the insulating nature is due to the flat band. We use a similar
  definition even in the presence of correlation as all of these phases are
  insulating and host edge modes with characteristics similar to the $U$=0
  case.}\BibitemShut {Stop}%
\end{thebibliography}
%-----------------------------------------------

\end{document}